\documentclass[a4paper,11pt]{article}
\usepackage{pos}
\usepackage{siunitx}
\usepackage{caption}
\usepackage{subcaption}

\title{Silicon vertex and tracking detector R\&D for CLIC}
\manuallySeparateAuthors
\author*[a, b]{Katharina Dort}
\author{ on behalf of the CLICdp Collaboration}

\affiliation[a]{CERN,\\
  Espl. des Particules 1, 1211 Meyrin, Switzerland}

\affiliation[b]{Experimental Physics II, University of Giessen,\\
Ludwigstr. 23, 35390 Giessen, Germany}

\emailAdd{katharina.dort@cern.ch}

\abstract{
	
	The physics aims at the proposed future high-energy linear e+e- collider CLIC pose challenging demands on the performance of the detector system. In particular, the vertex and tracking detectors have to combine a spatial resolution of a few micrometres and a low material budget with a time-stamping accuracy of a few nanoseconds. For the vertex detector, fine-pitch sensors, dedicated 65nm readout ASICs, fine-pitch bonding techniques using solder bumps or anisotropic conductive films as well as monolithic devices based on Silicon-On-Insulator technology are explored. Fully monolithic CMOS sensors with large  and small collection electrodes are under investigation for the large surface CLIC tracker. This contribution gives an overview of the CLIC vertex and tracking detector R\&D, focusing on recent results from test-beam campaigns and simulation-based sensor optimisation studies.
}

\FullConference{%
  40th International Conference on High Energy physics - ICHEP2020\\
  July 28 - August 6, 2020\\
  Prague, Czech Republic (virtual meeting)
}

\begin{document}
\maketitle
\section{Introduction}
The Compact Linear Collider (CLIC)~\cite{CLIC_2018_Summary} is a concept for a future linear electron-positron collider to be built at CERN. It is proposed to be constructed in three centre-of-mass energy stages ranging from 380\,GeV to 3\,GeV. 
The principle physics objectives of CLIC revolve around Standard Model top quark and precision Higgs physics as well as searches for Physics Beyond the Standard Model~\cite{CLIC_2018_Summary}. 

\section{Detector requirements}
\label{sec:experimental_conditions} 

The physics-driven requirements and experimental conditions at CLIC pose challenging requirements on the detector system~\cite{Dannheim:2673779}. 
A hit time tagging resolution of $\sim 5$\,ns is required to mitigate the impact of beam-induced background. 
Additionally, a single-plane spatial resolution of $< \SI{3}{\micro m}$ for the vertex detector and $< \SI{7}{\micro m}$ for the tracker, and a low mass constraint of $\sim 0.2 \% X_0$ per layer for the vertex and $\sim 1 - 2 \% X_0$ for the tracking detector are needed to comply with the required measurement accuracy. 
An average power dissipation of < 50 mW/cm$^2$ for the vertex is required for forced air cooling. 
The leak-less water cooling system foreseen for the tracker allows for an average power dissipation of < 150 mW/cm$^2$.
The hit detection efficiency for both vertex and tracking detector should exceed 99.5 \%.

\section{Hybrid Pixel Detectors}
A small pixel pitch of $\leq \SI{25}{\micro m}$ is required to limit the occupancy in the vertex detector to a few percent. 
Hybrid pixel detector technologies, separating the readout circuit from the sensitive volume, are capable of incorporating complex functionalities in a small pixel volume. 
Different technologies for readout ASICs and sensors are therefore under investigation for the CLIC vertex detector.

To this end, a 65\,nm CMOS process readout chip, the CLICpix2, has been designed with a pixel pitch of $\SI{25x25}{\micro m}$~\cite{santin2016clicpix2}.  
CLICpix2 ASICs have been solder bump-bonded to planar silicon sensors of various thicknesses ranging from $\SI{50}{\micro m}$ to $\SI{150}{\micro m}$.
The chip features an active matrix of 128\,x\,128 pixels and provides simultaneous 8-bit \textit{Time of Arrival (ToA)} and 5-bit \textit{Time over Threshold (ToT)}  measurement.

A single-chip bump-bonding process for CLICpix2 ASICs has been studies. 
The interconnect yield of the fine-pitch bump bonding was tested extensively in  laboratory measurements and a yield of up to 97.9\% was found~\cite{Williams:2706560} . 

Moreover, CLICpix2 assemblies have been tested in various test-beam campaigns and a spatial resolution of down to $\SI{3.2}{\micro m}$ was found for a sensor thickness of $\SI{130}{\micro m}$. 
For sensors sufficiently thin to comply with the vertex requirements on material budget, the spatial resolution degrades since charge sharing is not sufficient.
Therefore, innovative sensor concepts with enhanced charge sharing in the sensor layer are under investigation~\cite{Velyka:2019det}.   

\section{Monolithic Pixel Detectors}
 
In monolithic pixel detectors, the electronics circuits are integrated in the sensitive volume of the detector. 
This design allows to reduce the material budget and avoids the challenge of fine-pitch hybridization.  
These advantages and their large-scale production capabilities, make monolithic sensors  attractive candidates for the large-area CLIC tracker. 
However, as the electronics needs to be shielded from the high voltage applied to the sensor, it is placed in deep wells.
In \textit{High Voltage CMOS sensors (HV-CMOS)}, the circuitry is placed inside the collection electrode. 
With this design, high voltages can be applied. 
As a result, a large depletion zone and a high electric field are present in the sensing volume, which guarantee a fast charge collection via drift. 
However, the large collection electrode  leads to a comparatively high sensor capacitance.
It is beneficial to minimise the capacitance in order to profit from a lower power consumption, lower threshold and reduced noise level as well as an increased signal.  
In the \textit{High Resistivity CMOS (HR-CMOS)} design, the electronics is placed in wells separated from the collection electrode.
This allows to reduce the size of the collection electrode  leading to a reduced sensor capacitance. 

\subsection{High-Voltage CMOS sensors}

The ATLASpix\_Simple monolithic HV-CMOS test chip is produced in a commercial 180\,nm HV-CMOS process and features an active matrix of 25 columns and 400 rows of elongated pixels with a pitch of $\SI{130x40}{\micro m}$~\cite{Peric:2014faa}.
Each pixel records 10-bit ToA and 6-bit ToT information. 
The readout is realised in a  data-driven column-drain scheme. 

\begin{figure}[!tbp]
	\centering
	\begin{minipage}[b]{0.49\textwidth}
		\includegraphics[width=0.99\linewidth]{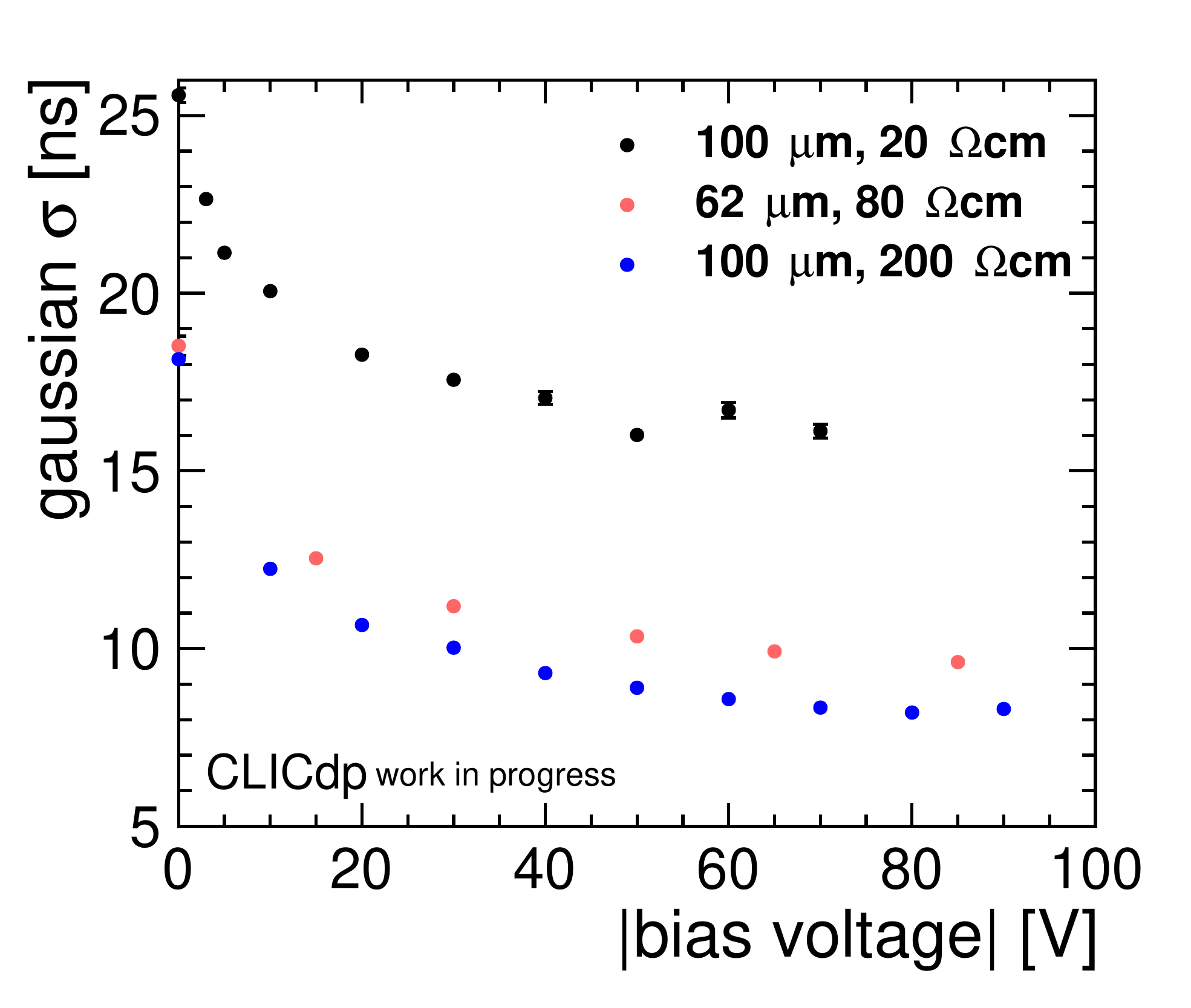}
		\caption{Timing resolution as a function of bias voltage for ATLASPix\_Simple assemblies with various resistivities.}
		\label{fig:atlaspix_spatial_resolution}
	\end{minipage}
	\hfill
	\begin{minipage}[b]{0.49\textwidth}
		\includegraphics[width=0.99\linewidth]{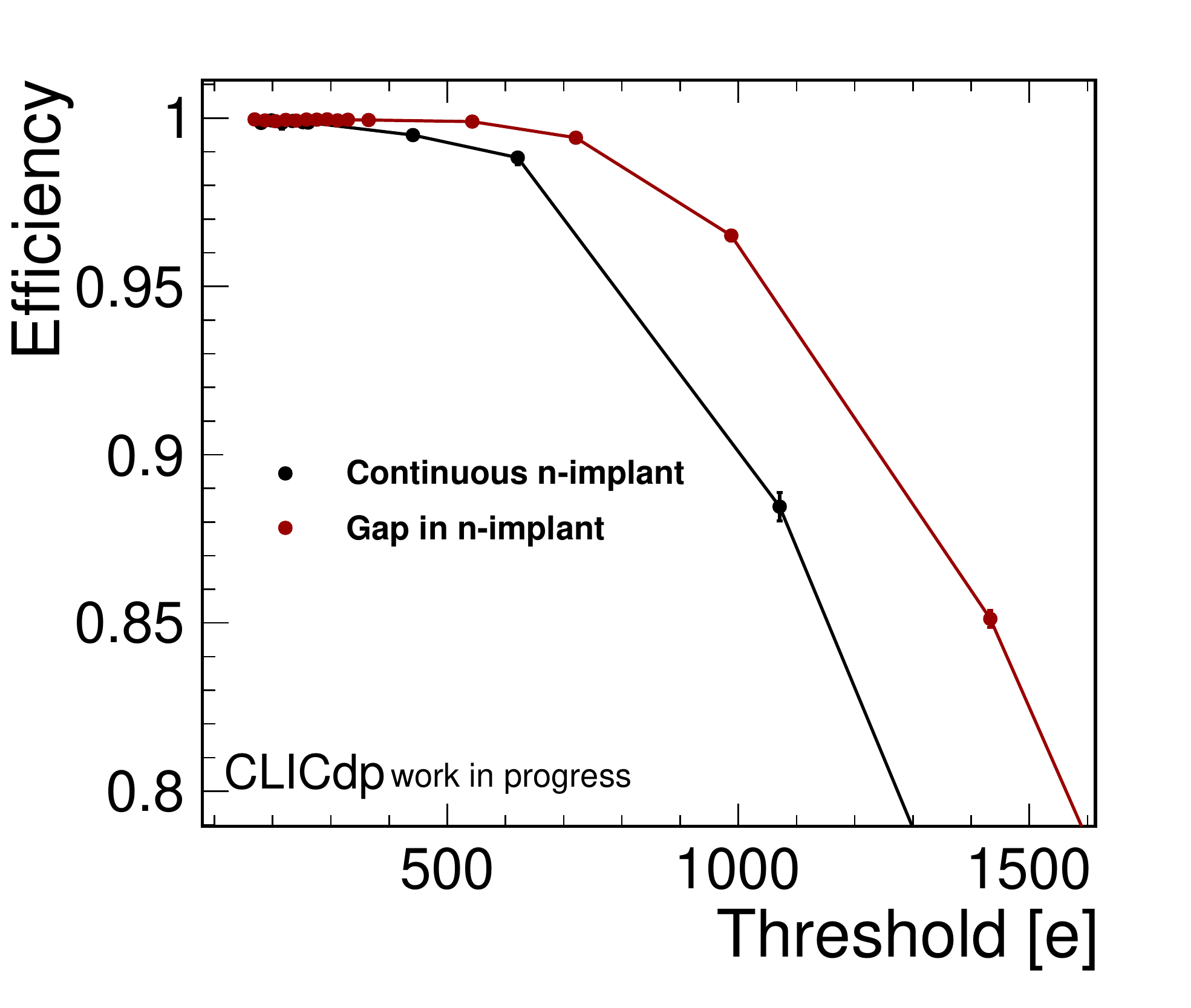}
		\caption{Hit detection efficiency as a function of detection threshold for the two CLICTD process variants with and without gap in the n-implant.}
		\label{fig:clictd_efficiency}
	\end{minipage}
\end{figure}

ATLASpix\_Simple  assemblies with wafer resistivities between $\SI{20}{\ohm cm}$ and $\SI{200}{\ohm cm}$ were studied in  test-beam campaigns at the DESY II test beam facility~\cite{Diener:2018qap}.
The timing resolution as a function of bias voltage for three different resistivities is depicted in Fig.~\ref{fig:atlaspix_spatial_resolution}
~\cite{Kroger:2020obh}.
The timing improves with increasing bias voltage and larger resistivity.
The higher bias and larger resistivity generate a larger depleted volume from which fast charge collection via drift is possible.
Moreover, a higher electric field is present in the sensor providing an additional acceleration in charge collection.

A hit detection efficiency well above 99.5 \% has been measured. 
While the efficiency decreases with increasing threshold, it has been shown that a larger resistivity and a higher bias voltage extend the efficient operation window~\cite{Kroger:2020obh}. 
The larger depleted volume, giving rise to a higher signal, is the reason for the better performance.

A spatial resolution of $~ \SI{11.3}{\micro m}$ has been measured in r$\Phi$-direction.
The spatial resolution is determined by the binary resolution (pitch / $\sqrt{12}$) since there is little charge sharing between neighbouring pixel cells. 
To improve the spatial resolution, the pixel layout has been adapted in a new test-chip based on the ATLASpix\_Simple design.
The chip is currently under investigation.

\subsection{High-Resistivity CMOS sensors}
The CLICTD chip is fabricated in a modified 180\,nm CMOS imaging process. 
It features a  matrix of 16\,x\,128 readout channels with a pitch of $\SI{300x30}{\micro m}$.
Each channel is segmented into eight  $\SI{37.5x30}{\micro m}$ sub-pixels with a dedicated analogue front-end. 
The sub-pixel comparator outputs are combined through a logical \textit{OR} in the digital front-end of a channel, which provides an 8-bit ToA and 5-bit ToT measurement. The bit pattern of sub-pixel hits is stored as well.
This readout architecture allows to reduce the digital logic while maintaining prompt charge collection~\cite{clictd_design_characterization}.

The sensor consists of a  $\SI{30}{\micro m}$ high resistivity epitaxial layer in which full lateral depletion can be achieved owing to the introduction of an n-type implant below the collection diode~\cite{SNOEYS201790}.  
CLICTD was fabricated in two sensor process variants, one featuring a continuous n-implant and the second one with a gap in the n-implant. 
The gap in the n-implant provides accelerated charge collection leading to an improved timing resolution and reduced charge sharing~\cite{Munker_2019}. 
As charge sharing is desired in the r$\Phi$-dimension of the detector to improve spatial resolution, the gap is only introduced in the perpendicular dimension, which corresponds to the direction along the $\SI{300}{\micro m}$ pitch. 


In test-beam campaigns at the DESY II test-beam facility, the spatial resolution of the detector in the r$\Phi$-dimension was found to be $\SI{4.6}{\micro m}$ and the timing resolution $\SI{5.8}{ns}$ for the design with continuous n-layer.  
The efficiency of the chip is $> 99\% $ up to a threshold of $\sim 450$\,e, as illustrated in Fig.~\ref{fig:clictd_efficiency}. 
The loss in efficiency for higher threshold values is less severe for the process with gap in the n-implant owing to the reduced charge sharing, which gives rise to a higher seed signal.
The larger efficient operation window is essential for future detectors in 65\,nm CMOS design in order to compensate for the reduced amount of charge generated in the thin sensors required for this technology.

The sensor design of CLICTD was optimised in 3D TCAD simulations.
For verification, the response of the sensor to a particle beam was studied in a combination of Monte Carlo (MC) and electrostatic TCAD simulations.
For this purpose, electrostatic sensor simulations in 3D TCAD were imported into the Geant4-based MC framework Allpix$^2$~\cite{apsq}, thereby harnessing the full potential of the accurate sensor modelling in TCAD and the high statistics thanks to the MC approach~\cite{allpix-hrcmos}. 


\begin{figure}
	\centering
	\begin{subfigure}{.5\textwidth}
		\centering
		\includegraphics[width=.99\linewidth]{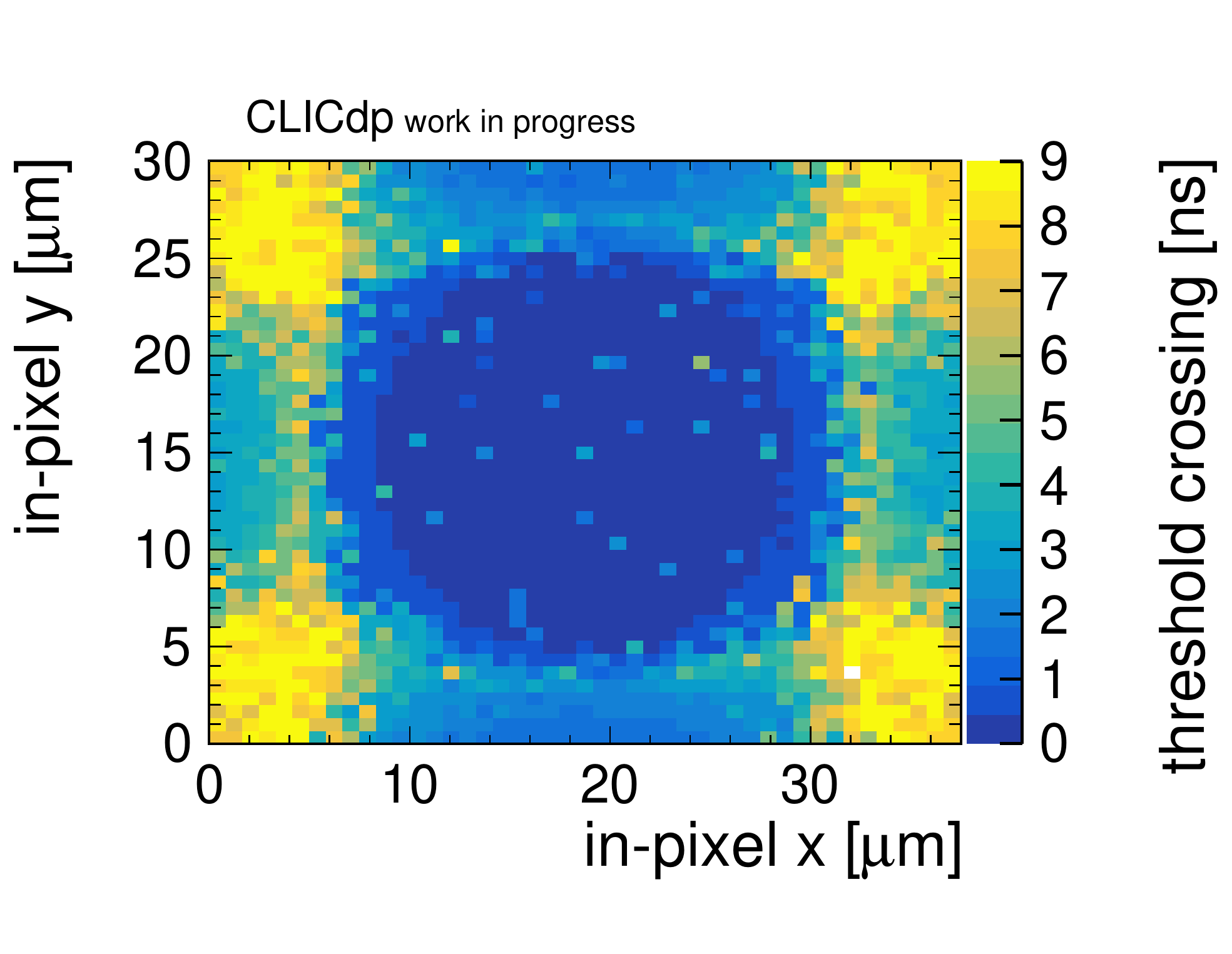}
	\end{subfigure}%
	\begin{subfigure}{.5\textwidth}
		\centering
		\includegraphics[width=.99\linewidth]{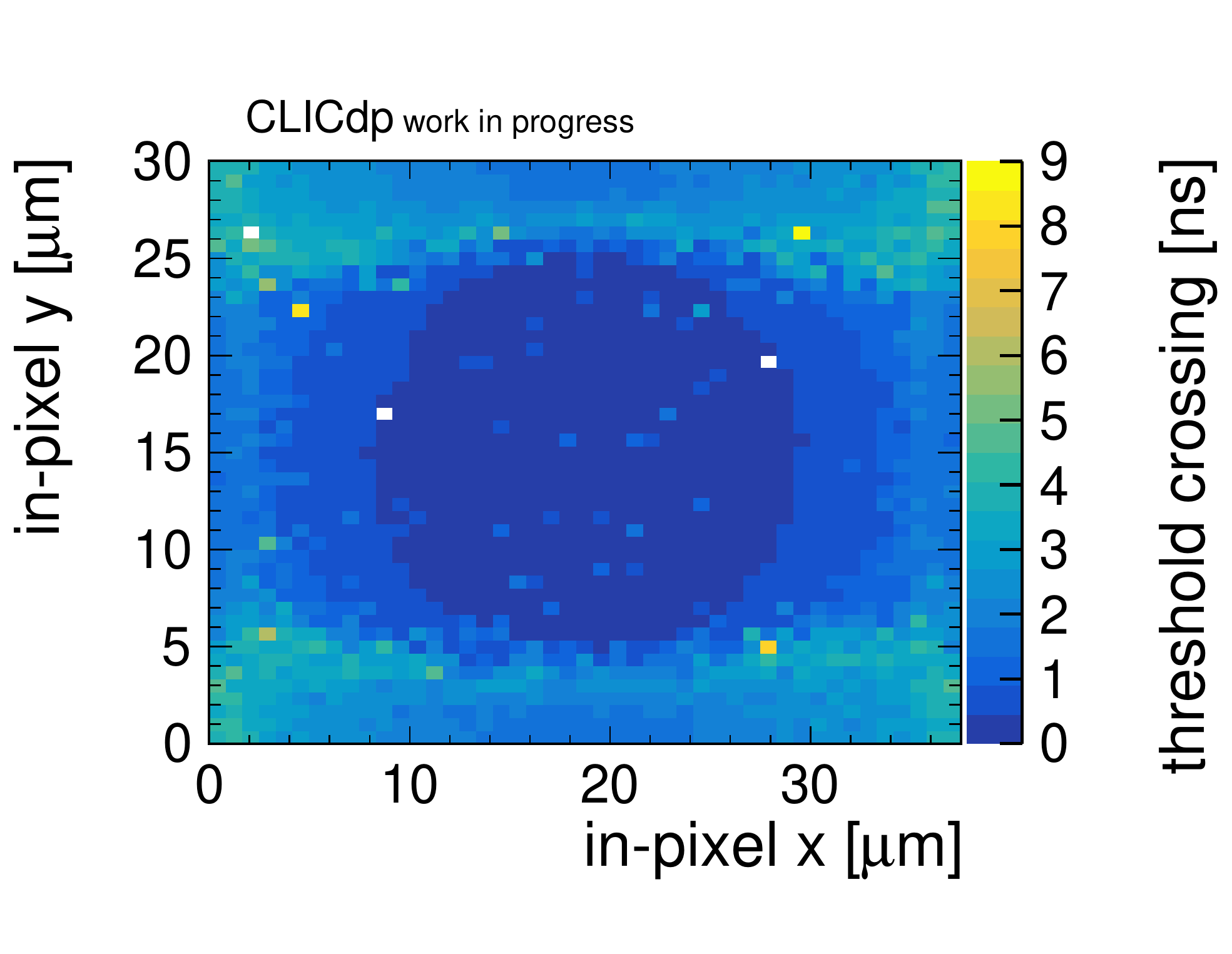}
	\end{subfigure}
	\caption{Simulated in-pixel threshold crossing time for the CLICTD process with continuous n-implant (left) and gap in the implant (right).}
	\label{fig:clictd_inpixel_timing}
\end{figure}

The threshold crossing time of the simulated pulses within a pixel cell is depicted in Fig.~\ref{fig:clictd_inpixel_timing} for both process variants. 
For the process without the gap, the sensor timing resolution degrades towards the pixel edges and particularly towards the pixel corners due to the increased charge collection time. 
The gap, introduced at very low and high x-values in the pixel cell shown on the right of Fig.~\ref{fig:clictd_inpixel_timing}, compensates for this sensor timing degradation by providing accelerated charge collection along the x-dimension. 
As a result, the threshold crossing time is improved and more homogeneous across the pixel cell.  
In the test-beam measurements, the timing resolution of CLICTD is limited by the sensor front-end circuits. 
Differences in timing resolution between the two process variants are therefore overshadowed by the front-end response.

The average power consumption for CLICTD has been estimated from simulations and static power consumption measurements. 
For the periphery, the average power consumption is 49\,mW and for the pixel matrix 5.7\,mW/cm$^2$, which is well in line with the CLIC tracking detector requirements~\cite{clictd_design_characterization}.

\section{Summary and Outlook}
A broad R\&D programme is being pursued to meet the demanding requirements imposed on the vertex and tracking detectors for CLIC.
For the research on hybrid pixel detectors, novel ASIC and sensor designs are developed and innovative interconnection technologies are explored. 
Both small and large collection electrode monolithic CMOS sensors are studied for the CLIC tracker. 
The HV-CMOS test chip ATLASpix\_Simple has a timing resolution of $\SI{6.8}{ns}$ and a hit detection efficiency of > 99.5\% which fulfil the CLIC tracker requirements. However, the spatial resolution of $\SI{11.3}{\micro m}$ is not compatible with the requirements.
Therefore, a new test chip with modified pixel layout to improve the spatial resolution has been produced and is currently under investigation.

Test-beam results of the CLICTD HR-CMOS sensor have shown that a spatial resolution of $\SI{4.6}{\micro m}$, a timing resolution of $\SI{5.8}{ns}$ and full efficiency can be achieved, which meet the tracking detector requirements.
Additionally, the estimated power consumption does not exceed the limit set for the CLIC tracker. 
The performance of CLICTD for inclined particle tracks is under investigation. 

\acknowledgments

This work has been sponsored by the Wolfgang Gentner Programme of the German Federal Ministry of Education and Research (grant no. 05E15CHA).
We would like to gratefully acknowledge CERN and their accelerator staff for the reliable test-beam operation.
Part of the measurements leading to these results have been performed at the Test Beam Facility at DESY II Hamburg (Germany), a member of the Helmholtz Association (HGF).
This project has received funding from the European Union’s Horizon 2020 research and innovation programme under grant agreement No 654168.

\bibliographystyle{JHEP}	
\bibliography{bibliography}


\end{document}